\documentstyle[aps,amssymb,pre,multicol,psfig]{revtex} 
\begin{document}

\draft    
\author{N.V.  Alexeeva \cite{add:nora} \cite{email:nora}, 
I.V.  Barashenkov
\cite{add:nora} \cite{email:nora}} \address{Max-Planck-Institut f\"ur Physik
komplexer
Systeme, N\"othnitzer Str.38, Dresden, Germany}
\author {G.P. Tsironis 
\cite{email:george}} \address{Physics Department, University of Crete
and FORTH, P.O.  Box
2208, 71003 Heraklion, Crete, Greece}

\title{Impurity-induced stabilization
of solitons in arrays of \\parametrically driven 
nonlinear oscillators}

\maketitle 
\begin{abstract}
Chains of parametrically driven, damped pendula
are known to support soliton-like clusters of in-phase
motion which 
become unstable and seed spatiotemporal chaos
 for sufficiently large driving amplitudes.
We show that the pinning of the soliton 
on a ``long" impurity  (a longer pendulum) expands dramatically its stability region 
whereas ``short" defects simply repel solitons producing 
effective partition of the chain. We also show that defects may spontaneously
nucleate solitons.
\end{abstract}


\begin{multicols}{2}
One of the most exciting 
 developments in  studies of 
  collective synchronization 
of arrays of coupled nonlinear oscillators 
 was the discovery of a dramatic
 increase of the degree of synchronization achieved 
  by disordering the array
\cite{Braiman-1,Braiman-2,Osipov}.  
This raised hopes of the possibility of
  ``taming spatiotemporal chaos with disorder". 
In an attempt to understand the nature of this phenomenon,
 a numerical study of the effect of a single impurity on 
an otherwise homogeneous chain was carried out
\cite{George}. Surprisingly, it turned out that a single impurity
  can produce simple spatiotemporal patterns
  in place of complex chaotic behaviour for very long chains of
  oscillators. 
   
  The  system simulated in Ref.\cite{George} (see also \cite{Braiman-2})
  consisted of a  chain of  pendula coupled to their nearest
 neighbours and driven by a periodic torque. 
All individual pendula in the array were in their chaotic parameter 
regime while the natural frequency of 
the central pendulum (the impurity) was out of the chaotic range.
Consequently, the suggested mechanism of stabilization was the 
formation of a nonchaotic cluster around the defect
which would subsequently pull the whole chain out of chaos \cite{Braiman-2,George}.  
  
   In the present note we study the effect of an
  impurity on a damped driven system by
 considering it from the viewpoint of nonlinear waves;
 that is,  we explore {\it collective\/} 
   stabilization mechanisms.  
   As in Ref.\cite{George} we  study a chain of  pendula but 
 unlike \cite{George}, 
   we focus on the regime where 
  all individual pendula are nonchaotic. (In spite of this fact
  the array as a whole may  fall into the state of spatiotemporal
  chaos.) Another distinction from Ref.\cite{George} is that our chain is driven 
  parametrically not externally; the reason is the 
availability of explicit solutions.

 We  assume that our pendula are strongly coupled.
  As a result, they tend to form soliton-like 
  clusters of coherent behaviour.
 Similarly to  externally driven systems, stationary
 solitons in parametrically driven chains are known to be stable 
 for small driving strengths but lose their stability to oscillating solitons
 as the driver's amplitude is increased \cite{BBK,Bond,Alex}. 
 Increasing the driving amplitude still further, a spatiotemporal
 chaotic state sets in  --- with or without a series of intermediate 
 bifurcations \cite{Bond}.
 
 We will demonstrate that  ``long" impurities may act as centers
 of {\it spontaneous\/} nucleation of solitons and hence
 in  chains with impurities solitons are even more
 generic structures than in homogeneous arrays.
 We 
 will prove that
   pinning of a stationary 
 soliton on a ``long"
 impurity expands its region of stability. 
In particular, by choosing a sufficiently long impurity pendulum,
 the soliton can be stabilized in the parameter region where 
 in the absence of the
  inhomogeneity it would set off the spatiotemporal chaos.
  Finally,
  although solitons pinned on ``short" impurities
  will turn out to  be more exposed to oscillatory instabilities,
  we will show that such a pinning is an unlikely occurrence
  due to the repulsion between solitons and the short defects.
  
The angle of the $n$-th pendulum  in our chain satisfies
\begin{eqnarray}
\label{chain1}
m l_n^2 {\ddot \theta_n} + \alpha l_n {\dot \theta_n} 
- k (\theta_{n+1} -2 \theta_n +\theta_{n-1})= 
\nonumber \\
=- m l_n \left( g + 4\omega^2 \rho \cos{2 \omega t} \right) \sin{\theta_n},
\end{eqnarray}
where 
$k$ is the torsion-spring constant,
 $\alpha$ 
  the  friction in the pivots,  
$\rho$ and $\omega$ are the amplitude and  
 frequency of the driver, and the length $l_n=1$ for all $n \neq 0$.
In what 
follows we set $g=m=1$. Defining the spacing between
the pendula by $a=1/ \sqrt{k}$, the continuous limit of eq.(\ref{chain1})
is given by the  damped-driven sine-Gordon equation:
\begin{eqnarray}
[1 +{\tilde p} \delta (z)] \theta_{\tau \tau} 
 -\theta_{zz}  +
[1 +{\tilde q} \delta (z)]
\times \nonumber \\ \times \ [\alpha \theta_{\tau}+
(1+4 \omega^2 \rho \cos{2 \omega \tau}) \sin{\theta}]=0.
\label{sg1}
\end{eqnarray}
Here $\delta (z)$ is Dirac's $\delta$;
${\tilde p}=a (l_0^2-1)$ and ${\tilde q}=a (l_0 -1)$. 
Choosing the length of the central pendulum to be $l_0=1 \pm b 
\varepsilon$
with $\varepsilon \ll 1$,
 gives  ${\tilde p}=2 {\tilde q}=
4q \varepsilon$,
 where $q=\pm \frac12 a b $.
 
If the driving frequency is just below the edge of the continuous
spectrum of linear waves: $\omega^2=1-\varepsilon^2$, and assuming
that $\alpha$ and $\rho$ are small:
$\alpha=\varepsilon^2 \gamma, \quad 
4 \omega^2 \rho = 2 h \varepsilon^2$, Eq.(\ref{sg1}) has  
small-amplitude solutions of the form
\[
\theta=2\varepsilon \psi(T,X) e^{-i \omega \tau}
+ c.c. +  {\rm O}(\varepsilon^3),
\]
where $T= \varepsilon^2 \tau /2$, $X=\varepsilon z$ and
 $\psi$  satisfies 
the parametrically driven damped nonlinear Schr\"odinger  equation
\begin{equation} 
\label{impNLS}
i \psi_T +\psi_{XX} +2 |\psi|^2 \psi -\psi + 2q \delta
(X) \psi = h \psi^* -i \gamma \psi.
\end{equation}
\begin{figure}
\begin{center}
\psfig{file=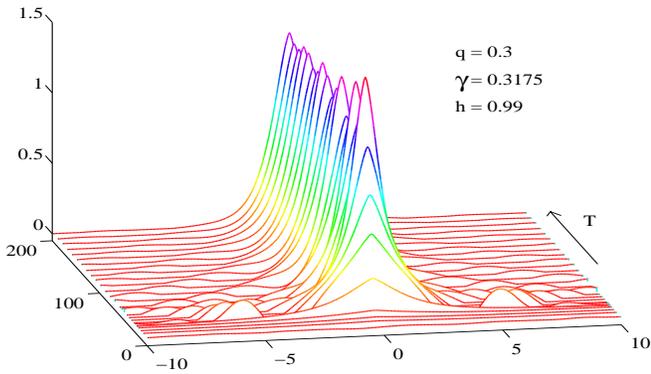,height=5cm,width=1.\linewidth}
\end{center}
\vspace{-1mm}
\caption{A  stable pinned soliton spontaneously created from
 a small-amplitude random
spatial distribution.}
\label{birth_from_noise}
\end{figure}
\noindent
This  nonlinear Schr\"odinger (NLS)
equation  was previously used to model 
the nonlinear Faraday
resonance in a long narrow water trough \cite{water}.
  The inhomogeneous term 
$2q \delta (X) \psi$ 
represents a local widening (for $q>0$) or narrowing ($q<0$) of the trough.
The same equation describes an easy-plane ferromagnet with a combination
of a static and  hf field in the easy plane 
\cite{magnetism,BBK} and a biaxial ferromagnet;
in both cases the inhomogeneous term accounts for an impurity spin.  
Equation (\ref{impNLS}) was also used in studies of
 the effect of phase-sensitive parametric
amplifiers on solitons in optical fibers \cite{optics}. 

As in the spatially 
homogeneous case $(q=0)$ \cite{BBK}, for $h>\sqrt{1+\gamma^2}$ the zero solution
$\psi=0$ is unstable against continuous spectrum excitations.
{\it ``Long"\/} impurities ($q >0$)   harbour a discrete mode 
 (here $\epsilon \ll 1$): 
\[
\delta \psi = \epsilon
\left( \cos \Omega T -i\frac{1+h-q^2}{\Omega}
\sin \Omega T
\right) e^{ -\gamma T-q|X|}, 
\]
$\Omega^2 =(1-q^2)^2-h^2$.
When $h> {\frak h}_{q, \gamma}= \sqrt{(1-q^2)^2 +\gamma^2}$, this localized mode also  produces instability.
For positive $q < \sqrt{2}$, in which case ${\frak h}_{q, \gamma}$ 
lies below $\sqrt{1+ \gamma^2}$,
  the nonlinear development of this instability  leads to the formation of
solitons (fig.\ref{birth_from_noise}), stable or unstable.

 Equation (\ref{impNLS}) exhibits
two stationary soliton solutions, each having
a cusp at the origin: 
\begin{eqnarray}
\psi_{\pm}(X)=A_{\pm} {\rm sech} \left( A_{\pm} |X|+{\tilde x} \right) 
e^{-i \theta_{\pm}}, 
\label{cuspsol}
\end{eqnarray}
where
$\cos{2 \theta_{\pm}}=\pm \sqrt{1-\gamma^2/h^2 }$, 
 $A_{\pm}^2= 1 \pm \sqrt{h^2- \gamma^2}$,  
${\tilde x}= {\rm arctanh}\left({q/A_{\pm}}\right)$.
 For weak impurities, $|q|<1$, the $\psi_+$
soliton exists for any $h$ and $\gamma$ 
satisfying $h > \gamma$,
whereas  the $\psi_-$ requires, in addition, that
$h< {\frak h}_{q,\gamma}$.
 Strong impurities ($|q|>1$) do 
not support the  $\psi_-$ soliton at all whereas the $\psi_+$ 
exists only if $h>{\frak h}_{q,\gamma}$.

First we  demonstrate
that the soliton  $\psi_-$ is  unstable 
and  can always be disregarded. 
Taking the linear perturbation
in the form $\delta \psi_{\pm} = A_{\pm}(f + i g) e^{-i \theta_{\pm}
- \gamma T}$ 
gives 
\begin{eqnarray} 
\label{linear}
-g_t-\Gamma g =L_1 f,
\quad
f_t - \Gamma f = L_0 g,
\end{eqnarray}
where  $ \Gamma=\gamma/A_{\pm}^2$, $t=A_{\pm}^2 T$, and the operators
\begin{eqnarray} 
\label{oper}
L_1 = -\partial_x^2+ 1 - 6 {\rm sech}^2 (|x|+{\tilde x}) - 2Q \delta (x), \\ 
L_0= - \partial_x^2 + 1 \mp 2H - 2 {\rm sech}^2 (|x|+{\tilde x}) -2Q \delta (x), 
\end{eqnarray}
with $Q=q/A_{\pm}=q\sqrt{1 \mp H}$, 
$H=\sqrt{h^2-\gamma^2}/A_{\pm}^2$, and 
 $x=A_{\pm}X$. 
The minimum eigenvalue of the operator $L_0$ associated with a nodeless eigenfunction 
${\rm sech\/} (|x|+{\tilde x})$, is  $\nu_0= \mp 2H$. 
Consequently, in the case of the $\psi_-$ soliton  $L_0$ is
positive definite and Eqs.(\ref{linear}) can be rewritten as 
\begin{equation}
L_0^{-1} f_{tt} = \Gamma^2 L_0^{-1}f -L_1 f.
\label{secorder}
\end{equation}
The   maximum exponential growth rate $\lambda$ of
solutions to Eq.(\ref{secorder}) is given by \cite{max}
\begin{eqnarray}
\lambda^2
=\Gamma^2+
 \sup_{f}
\frac{< f(x)|  - L_1| f(x)>}{< f(x) |L_0^{-1}| f(x) >}.
\label{sup}
\end{eqnarray}
For any $Q$ the operator $L_1$ has a negative eigenvalue
$\mu_0=1-\kappa^2$ associated with an even eigenfunction
\begin{equation}
y_0(x)=e^{-\kappa \xi} (3 \tanh^2\xi + 3 \kappa \tanh \xi
+ \kappa^2-1),
\label{y}
\end{equation}
where $\xi= |x|+{\tilde x}$ and $\kappa>1$ is a root of
$
\kappa^3 + 2 \kappa^2 Q + \kappa (3Q^2-5Q-4) + 3Q^3=0
$.
Thus    the  supremum
in (\ref{sup}) is positive, $\lambda$ is $> \Gamma$ and  the soliton
 $\psi_-$
is  unstable against a symmetric nonoscillatory mode for all $q$, $h$ and $\gamma$.

A similar argument can be used to detect the 
disengagement instability of  the other soliton,
$\psi_+$, arising for ``short" impurities ($q<0$). We simply
notice that the second lowest eigenvalue of $L_0$ which is
associated with an odd eigenfunction
\[
w_1(x)= \mbox{sgn} (x) e^{Q(|x| + {\tilde x}) }
\left\{ \mbox{tanh} (|x|+ {\tilde x}) -Q \right\},
\]
is equal to $\nu_1=1-2H-Q^2$. For
$2H< 1-Q^2$ or equivalently for $h< {\frak h}_{q,\gamma}$,
the operator $L_0$ is positive definite on the subspace of odd
functions. 
On the other hand,
for $Q<0$ the operator $L_1$ has
 a negative eigenvalue $\mu_1=1-\kappa^2$
associated with an odd eigenfunction $y_1(x)= {\rm sgn\/}x \ y_0(x)$,
with $y_0$ as in (\ref{y}) and $\kappa>1$ a root of
$\kappa^2 + 3 \kappa Q + 3 Q^2-1=0$. 
Hence the variational principle (\ref{sup}) is still 
applicable, $\lambda$ is $> \Gamma$ and the $\psi_+$ is
unstable.
 The interpretation of this instability is straightforward
if one notices that in the conservative case ($\gamma=0$) the inhomogeneous
term $-2q\delta(X) |\psi|^2$ produces a local decrease respectively
increase of energy
for $q>0$ respectively $q<0$. Consequently, in the conservative and weakly dissipative
cases, the ``long" impurity will  attract and the ``short" one 
repel  small-amplitude tails
of  {\it distant\/} solitons.
On the other hand, the energy of the {\it pinned\/} soliton 
is $E_Q=\frac43 A^3 (1-3Q+2Q^3)$. For $Q>0$ ($Q<0$) this
is smaller
(greater) than the energy  $E_0$ of the infinitely remote soliton.
 These two facts indicate that  $Q>0$-impurities 
should attract and trap solitons
(cf.\cite{BKG}). In the $Q<0$ case,
conversely, distant solitons will be repelled  while an initially pinned 
  soliton   will depin and move away from the impurity regaining its
  cusp-free shape.
  \newpage
\vspace{-40mm}
\begin{figure}
\begin{center}
\psfig{file=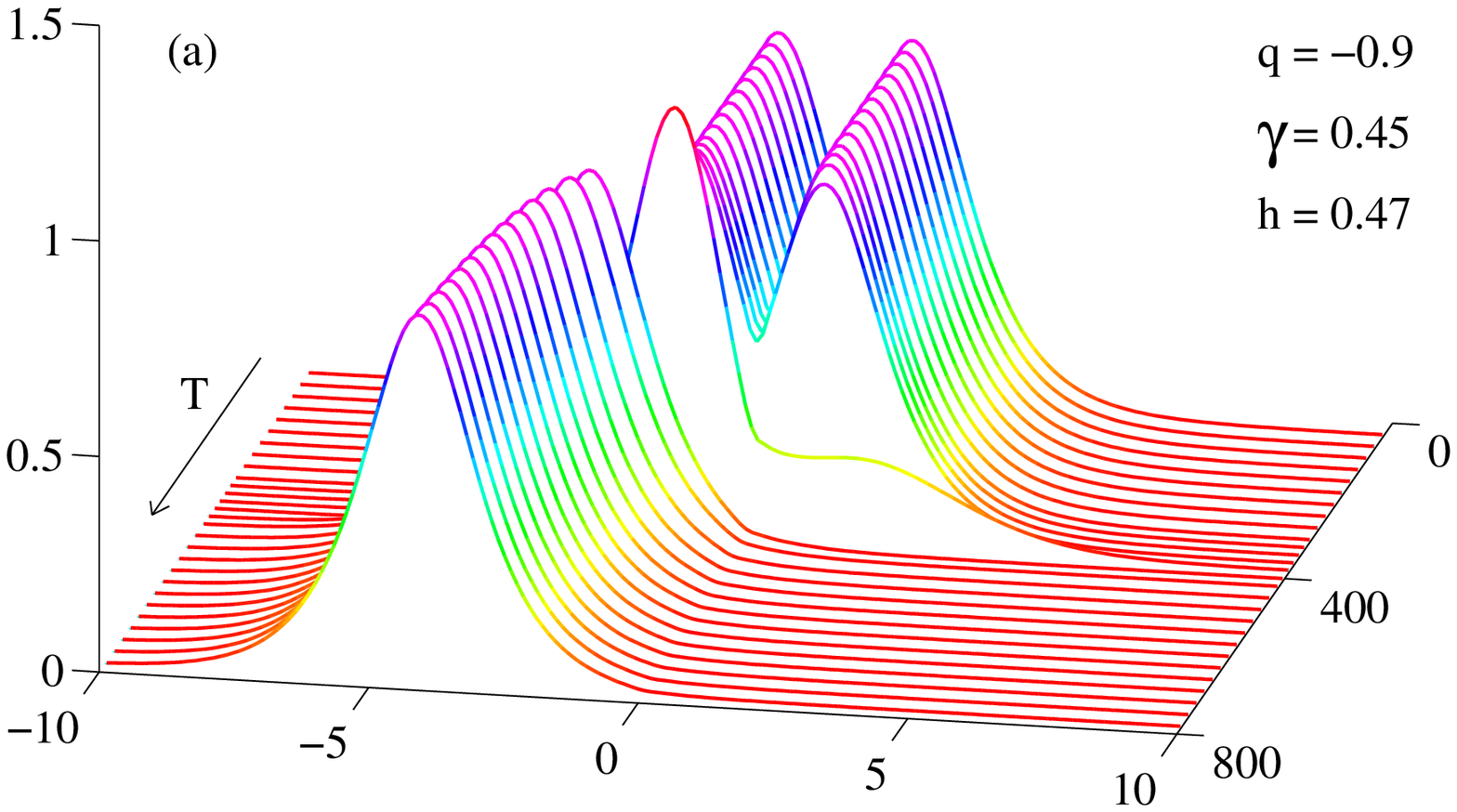,height=5cm,width=1.\linewidth}
\psfig{file=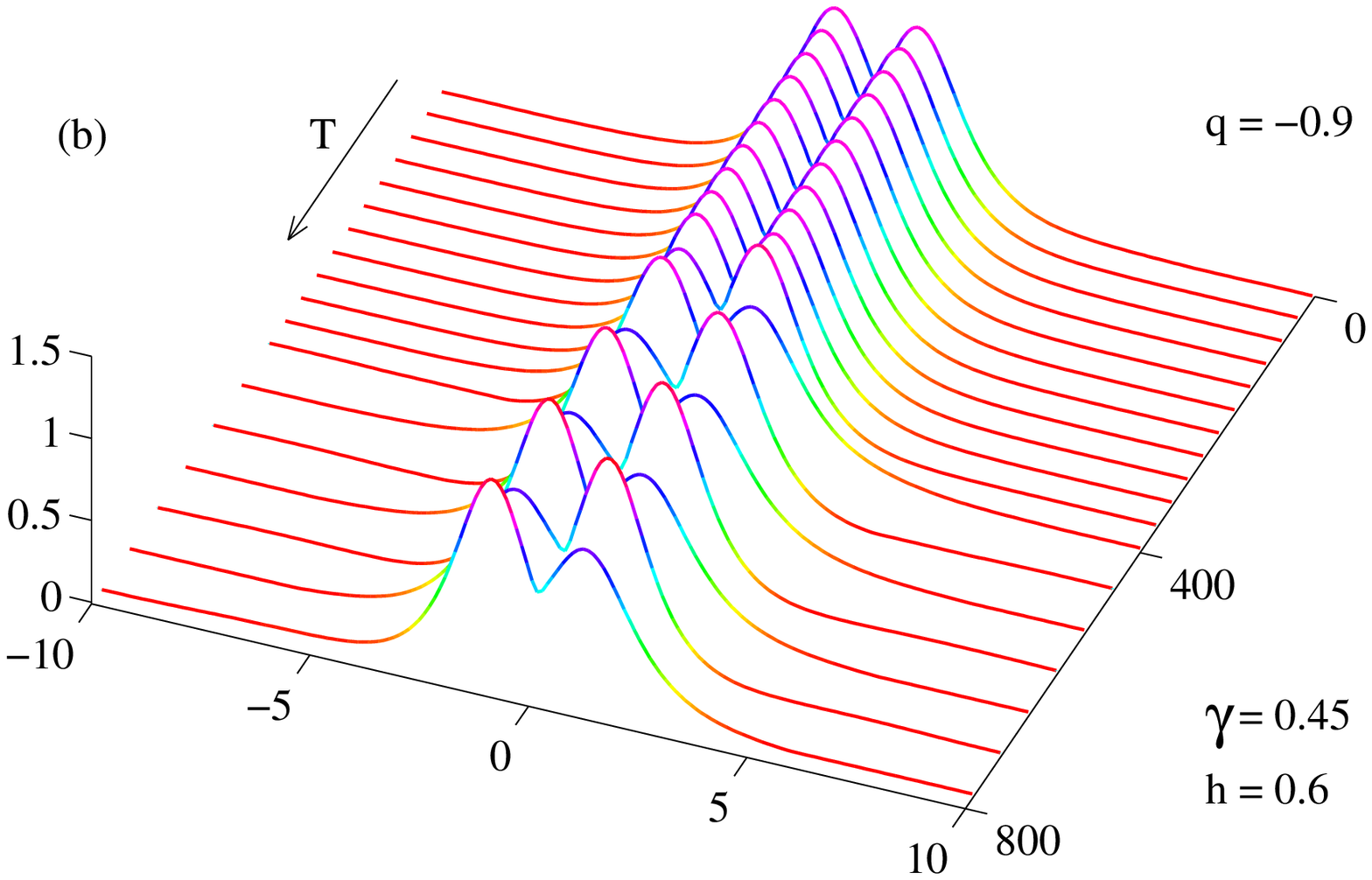,height=5cm,width=1.\linewidth}
\end{center}
\vspace{-3mm}
\caption{The effect of 
the $q<0$ impurity.
For $h< {\frak h}_{q,\gamma}$, the growth of the
asymmetric instability leads to the
soliton's depinning and repulsion from the impurity (a),
while
for $h>{\frak h}_{q,\gamma}$, the 
asymmetric Hopf instability evolves into a two-humped soliton
staggering in a seesaw fashion (b).}
\label{asymmetric}
\end{figure}
\noindent
This was indeed confirmed by simulations of Eq.(\ref{impNLS})
 (fig.\ref{asymmetric}(a)).
  It is fitting to note that the depinning  instability
is not  connected with overdriving the chain; it
occurs  already in the undriven NLS  \cite{BKG}.

In the region $h> {\frak h}_{q,\gamma}$
(as well  as in the case of symmetric
  instabilities, and  for long impurities $q>0$) the variational principle (\ref{sup}) is not
  applicable. 
  Here
we let 
   $f(x,t)=u(x) e^{i \Omega t}$ and 
$g(x,t)= - \frac{\omega }{\Gamma + i \Omega} v(x) e^{i \Omega t}$,
where $\Omega^2= \omega^2- \Gamma^2$.  Eq.(\ref{linear}) reduces to
 an eigenvalue 
problem 
\begin{equation}
L_1u = \omega v, \quad 
L_0 v= \omega u,
\label{EV}
\end{equation} 
which we solved numerically.
Notice that we have reduced a three- to two-parametric problem.
Having found $\omega = \omega (q,H)$, one immediately recovers the instability
growth rate $|{\rm Im} \, \Omega(q,H)|- \Gamma$ for all $q$,  $h$, and $\gamma$.

 For small $h< {\frak h}_{q, \gamma}$
there is only one unstable pair of  imaginary eigenvalues $\pm \omega_1$,
with the  associated  
 $u$ and $v$ being odd. As $h$ is increased beyond ${\frak h}_{q,\gamma}$,
the imaginary eigenvalues  move onto 
the real axis (i.e. the soliton restabilizes) while another real doublet $\pm \omega_2$
detaches from the continuous spectrum. The two collide and emerge as a complex quadruplet
after which the real and imaginary parts grow until an asymmetic instability
sets in. This is now {\it not\/} a disengagement instability;
the stationary soliton is replaced by a two-humped structure
(still pinned  
\begin{figure}
\begin{center}
\psfig{file=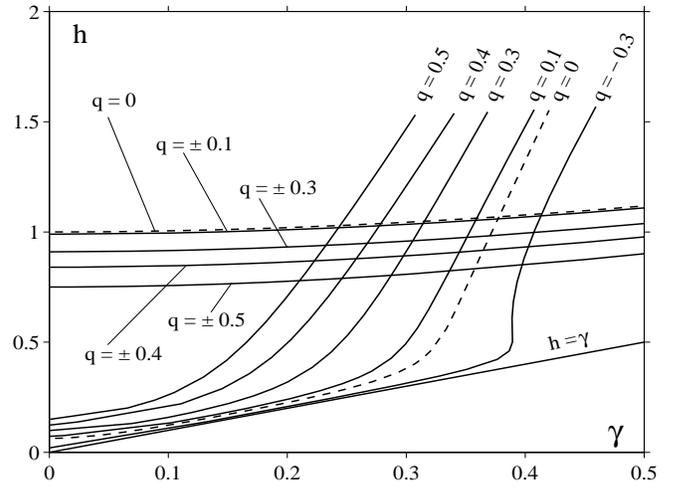,height=6.5cm,width=1.\linewidth}
\end{center}
\vspace{-10mm}
\caption{Atlas of stability charts on the $(\gamma,h)$-plane.
Above the dashed ``horizontal" curve,
$h=\sqrt{1+ \gamma^2}$, all localized solutions are unstable
 w.r.t. continuous spectrum waves.  
  Solid ``horizontal" curves are
$h={\frak h}_{q, \gamma}$.  
Below their 
corresponding
 ${\frak h}_{q, \gamma}$-curves
impurities with $q<0$ repel solitons. 
The family of ``parabolas" depict the onset of (symmetric)
  instability of the pinned soliton; 
  the greater is $q$ the larger is the
stability domain. Finally,
in the region between $\sqrt{1+ \gamma^2}$ and
${\frak h}_{q, \gamma}$ a $q>0$-impurity
will  spontaneously nucleate solitons.  }
\label{chart}
\end{figure}  
\noindent
on the impurity) whose left and right 
wings oscillate $180^{\circ}$ out of phase (fig.\ref{asymmetric}(b)).

When $q>0$, the motion of eigenvalues $\omega$
on the complex plane
 is similar to
 the homogeneous case \cite{BBK}. The 
soliton $\psi_+$ is stable for $h$ close to $\gamma$ but loses its 
stability to a symmetric oscillating soliton as $h$ is increased. 
Fig.\ref{chart} shows the Hopf bifurcation curves $h=h_q(\gamma)$
obtained from the relation $|{\rm Im} \, \Omega(q,H)|= \Gamma$
for $q=0.1, \, 0.3, \, 0.4, \, 0.5$ and $-0.3$. For $q>0$ the
stability domain is wider than without an impurity. 
For example, in a chain with 
the coupling $k=0.5$ driven at $\omega \approx 0.98$, 
lengthening the central 
pendulum by  15\% (which gives $q=0.5$) is sufficient to {\it double\/}
  the
size of the soliton's
stability domain. If the same chain is driven at $\omega \approx
0.999$, the $q=0.5$ and hence the same effect is produced just by 
a 3\% elongation
of the central pendulum. On the contrary,
the  $q<0$ impurity narrows the stability domain.

\begin{figure}
\begin{center}
\psfig{file=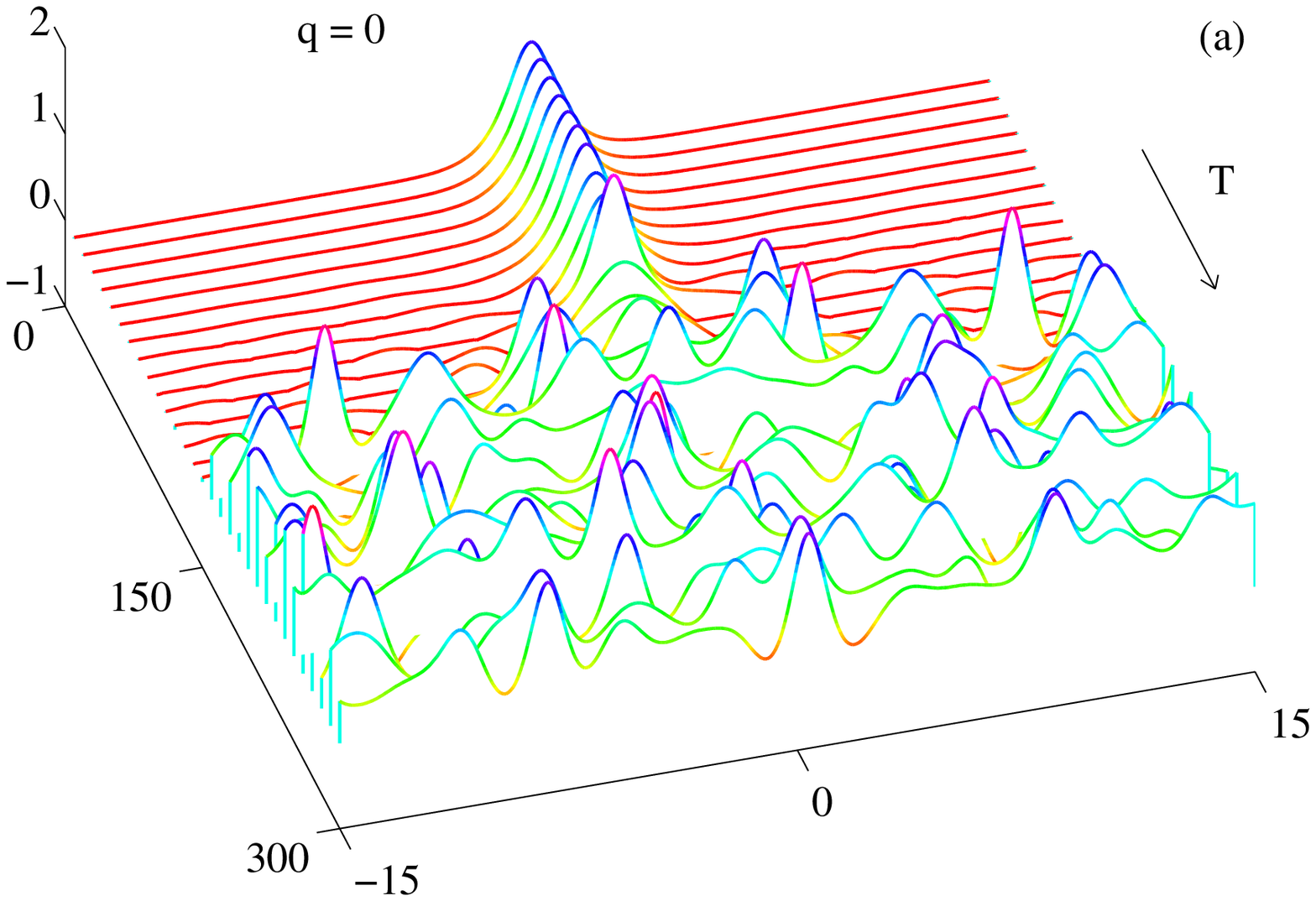,height=5cm,width=1.\linewidth}
\psfig{file=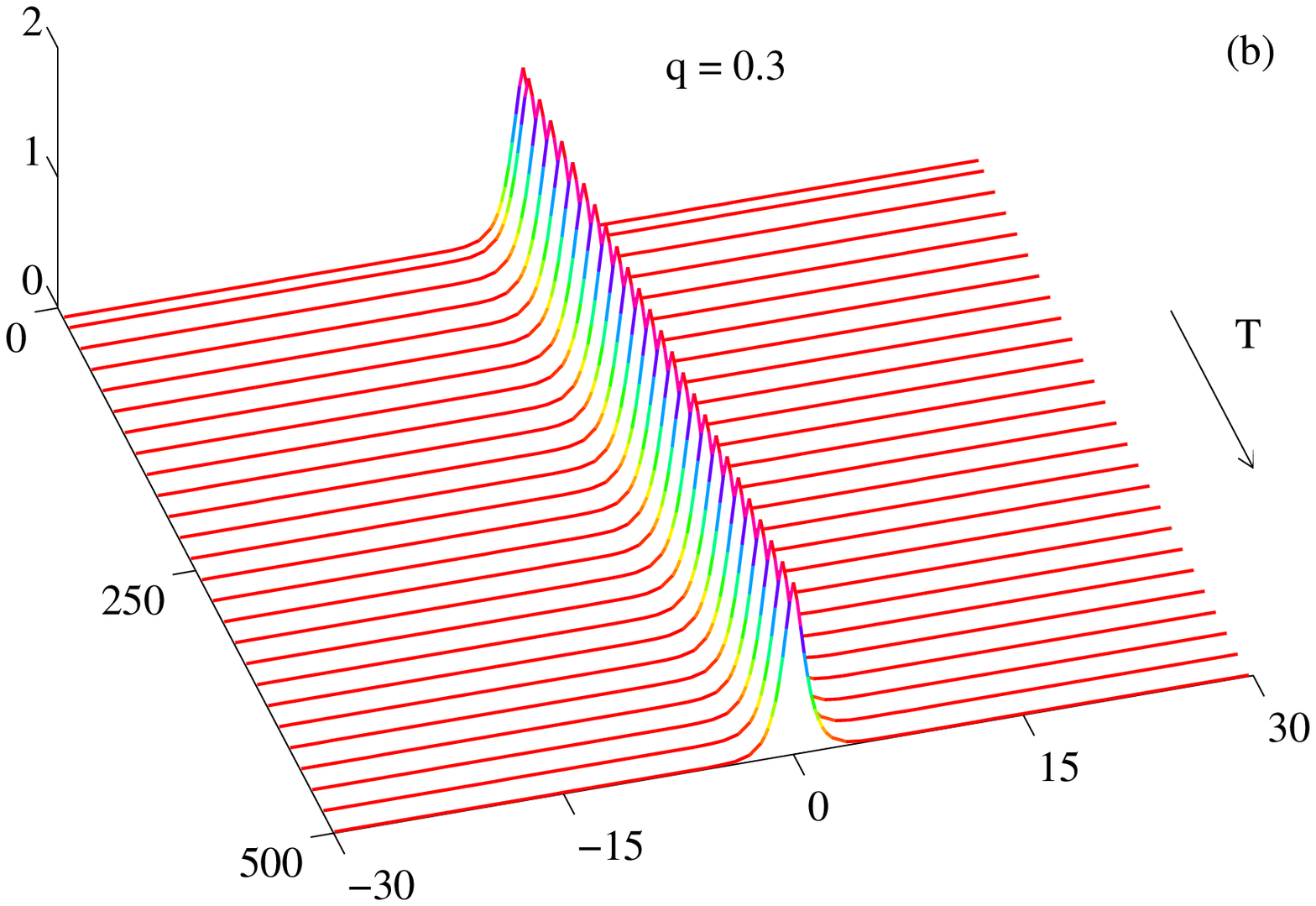,height=5cm,width=1.\linewidth}
\end{center}
\vspace{-5mm}
\caption{In the homogeneous chain 
with $\gamma=0.315, h=0.95$ the  unstable soliton 
seeds spatiotemporal chaos (a). 
Introducing the  impurity with $q=0.3$ is sufficient to
stabilize it (b).}
\label{tamed_chaos}
\end{figure}

 Thus,
{\it long\/} impurities exert a strong influence on solitons' dynamics.
For $h<{\frak h}_{q,\gamma}$ they   attract and trap
solitons;
for $h>{\frak h}_{q,\gamma}$ pinned solitons are spontaneously formed
around the $q>0$ defects. 
On the other hand,    
   the soliton with $h$ and $\gamma$ such that it
would ignite spatiotemporal chaos in the homogeneous case
 \cite{Bond}, is
stabilized when  pinned on a sufficiently long impurity
(fig.\ref{tamed_chaos}). Therefore the $q>0$ defects should have a stabilizing effect 
on the chain. One should keep in mind, however, that
spatiotemporal chaotic states are not localized and
\vspace{-15mm}
\noindent
a single 
stable soliton will clearly  be insufficient to
suppress chaos in a long chain. The chaos can always be triggered 
by choosing the initial
 condition  far enough from the soliton. In order to  suppress
chaos in a larger phase volume multiple impurities should be introduced; one is
therefore led to the necessity  of examining stability of solitons
and periodic waves on finite intervals. 

 Next,
the fact that {\it short\/} impurities  enhance the symmetric instability
should not play a destabilizing role  since  solitons tend to 
avoid ``short" defects. On the contrary, these repulsive inhomogeneities
will effectively partition the chain into smaller intervals
and this will generally have a stabilizing effect since 
long-wavelength instabilities will not fit in \cite{BS}.

\begin{figure}
\begin{center}
\psfig{file=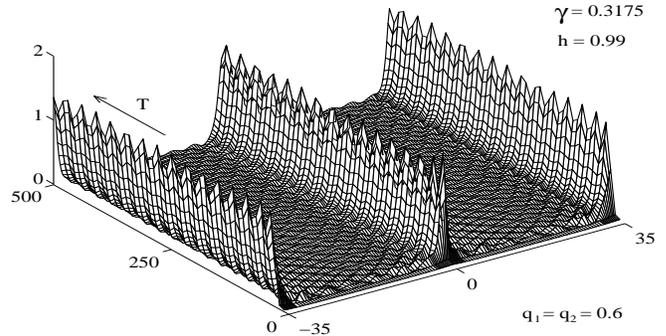,height=5cm,width=1.\linewidth}
\end{center}
\vspace{-3mm}
\caption{Two $q>0$ impurities  placed in the middle and
 at
the  end of the periodic
interval, respectively.}
\label{synchronisation}
\end{figure}

The analysis of the effect of multiple impurities (as well 
as the related case of finite intervals) is beyond the scope of this work.
Here we only mention  that  introducing  more than one impurity
can result in more complicated (though
still regular)  patterns. We illustrate this by simulating
the case of two impurities.
As one could expect, when two stable stationary  solitons are pinned
{\it very\/} 
far from each other, they do not interact and remain time-independent.
 However, if the separation is smaller than a certain
 critical distance,  they 
 start exchanging
weak radiation waves and
develope spontaneous oscillations which subsequently synchronize
(fig.\ref{synchronisation}.) 

 This research was 
supported by the FRD of South Africa, Max-Planck-Gesellschaft
and  FORTH of Greece.
\vspace{-6mm}
 

\end{multicols}
\end{document}